\documentclass[12pt]{article}%
\usepackage{amsmath,latexsym}
\usepackage{graphicx}
\usepackage{amsmath}
\usepackage{amsfonts}
\usepackage{amssymb}%
\begin{document}

\title{{Vacuum energy in a noncommutative-geometry
     setting}}
   \author{
Peter K.F. Kuhfittig*\\  \footnote{kuhfitti@msoe.edu}
 \small Department of Mathematics, Milwaukee School of
Engineering,\\
\small Milwaukee, Wisconsin 53202-3109, USA}

\date{}
 \maketitle

\begin{abstract}\noindent
It has recently been proposed that vacuum
energy is zero in spite of the quantum-field
fluctuations that occur everywhere, even at
absolute zero.  The implication is that dark
energy must have a different origin, unrelated
to vacuum energy.  The proposal is based on 
the use of a local Lorentz frame in a 
non-gravitational field; this also results in 
a stress-energy tensor of the vacuum that is a 
perfect fluid with equation of state 
$p_{\text{vac}}=-\rho_{\text{vac}}$.  It is 
noted in this paper that noncommutative 
geometry, an offshoot of string theory, yields 
the same equation of state without being 
confined to a local Lorentz frame.   As a 
result, the noncommutative-geometry background 
is able to account for the accelerated 
expansion and hence for dark energy.  So 
vacuum energy can only be zero if dark energy 
is indeed unrelated to vacuum energy. 


\end{abstract}

\section{Introduction}

In a recent paper, Gregory Ryskin \cite{gR20}
proposed that vacuum energy is zero, even
though the vacuum is filled with short-lived
virtual particle-antiparticle pairs due to
the fluctuations of quantum fields.  These
fluctuations, even at absolute zero, have
nonzero energy.

Ryskin uses a local Lorentz frame in a
non-gravitational field, as a result of which
the stress-energy tensor of the vacuum is
that of a perfect fluid with equation of
state
\begin{equation}\label{E:VAC}
   p_{\text{vac}}=-\rho_{\text{vac}}.
\end{equation}
According to Ref. \cite{gR20}, this equation
is a direct consequence of relativistic
invariance.  Moreover, the relativistic
invariance dictates the form of the
stress-energy tensor in terms of the
fundamental metric tensor:
$-\rho_{\text{vac}}g_{\mu\nu}$, where
$\rho_{\text{vac}}$ is the vacuum energy
density \cite{sW08}.  Since this tensor
vanishes in a particular frame of reference, 
it must vanish in all.  The point is that 
if the vacuum energy is indeed zero, then 
dark energy, first discovered in 1998, must
have a different origin, unrelated to
\emph{}vacuum energy.

It is proposed in this paper that given a
noncommutative-geometry background, we
arrive at the same equation of state, even
though we are no longer confined to a local
Lorentz frame.  This enables us to use
noncommutative geometry to account for the
accelerated expansion on both the local and
cosmological scales.  The two viewpoints
turn out to be consistent with each other.

\section{Noncommutative geometry}
Our first task is to review the basic ideas
in noncommutative geometry, an area that is
based on a certain outcome of string theory,
namely, that coordinates may become
noncommuting operators on a $D$-brane \cite
{eW96, SW99}.  This statement refers to the
commutator $[\textbf{x}^{\mu},\textbf{x}^{\nu}]
=i\theta^{\mu\nu}$, where $\theta^{\mu\nu}$
is an antisymmetric matrix.  Moreover,
noncommutativity replaces point-like
structures by smeared objects, thereby
elimination the divergences that normally
occur in general relativity \cite{SS03a,
SS03b, NS10}.  A natural way to accomplish
the smearing effect is to use a Gaussian
distribution of minimal length
$\sqrt{\beta}$ \cite {NSS06, pK13}.
Alternatively, one may assume that the energy
density of the static and spherically symmetric
and particle-like gravitational source has
the form \cite{NM08, pK15}
\begin{equation}\label{E:rho}
  \rho(r)=\frac{\mu\sqrt{\beta}}
     {\pi^2(r^2+\beta)^2}.
\end{equation}
This form can be interpreted to mean that
the mass $\mu$ of the particle is diffused
throughout the region of linear dimension
$\sqrt{\beta}$ due to the uncertainty.  Eq.
(\ref{E:rho}) leads to the mass distribution
\begin{equation}\label{E:mass}
   M(r)=\int^r_04\pi (r')^2\rho(r')\,dr'
  \\ = \frac{2M}{\pi}\left(\text{tan}^{-1}
  \frac{r}{\sqrt{\beta}}
  -\frac{r\sqrt{\beta}}{r^2+\beta}\right),
\end{equation}
where $M$ is now the total mass of the
source.

Returning now to Ref. \cite{NSS06}, we
examine the equation of state
(\ref{E:VAC}) from a different perspective.
Suppose we start with the stress-energy
tensor $T_{\alpha\beta}$ and the covariant
conservation equation $T^{\alpha}
_{\phantom{\beta r}\beta;\,\alpha}=0$.  If
$\beta =r$, we obtain
\begin{equation}
  \frac{\partial}{\partial r}T^r_{\phantom{tt}r}
  =-\frac{1}{2}g^{tt}\frac{\partial g_{tt}}
  {\partial r}
  (T^r_{\phantom{tt}r}-T^t_{\phantom{tt}t})-
  g^{\theta\theta}\frac{\partial g_{\theta\theta}}
  {\partial r}
  (T^r_{\phantom{tt}r}-
  T^\theta_{\phantom{tt}\theta})
\end{equation}
in terms of the stress-energy tensor and the
fundamental metric tensor.  According to Ref.
\cite{NSS06}, to preserve the property
$g_{tt}=-g^{-1}_{rr}$, we require that
$T^r_{\phantom{tt}r}=T^t_{\phantom{tt}t}=
-\rho(r)$, while
\begin{equation}
   T^\theta_{\phantom{tt}\theta}=-\rho(r)
   -\frac{r}{2}\frac{\partial\rho(r)}
   {\partial r}.
\end{equation}
The main conclusion is that a massive
structureless point is replaced by a
self-gravitating droplet of anisotropic
fluid of density $\rho$, yielding the
radial pressure
\begin{equation}\label{E:radial}
   p_r(r)=-\rho(r)
\end{equation}
and the tangential pressure
\begin{equation}\label{E:tangential}
    p_{\bot}(r)=-\rho(r)-\frac{r}{2}
    \frac{\partial\rho(r)}{\partial r}.
\end{equation}
It is also noted that on physical
grounds, the radial pressure is needed
to prevent the collapse to a matter point.

In seeking a connection to Eq. (\ref{E:VAC}),
we return to Ref. \cite{NSS06} once again:
it is emphasized that noncommutative geometry
is an intrinsic property of spacetime and
does not depend on any particular properties
such as curvature.  Moreover, the effects of
noncommutativity can be described as follows:
keep the standard form of the Einstein tensor
on the left-hand side of the field equations
and insert the modified stress-energy tensor
as a source on the right-hand side.  This
leads to the important conclusion that the
length scales need not be microscopic.  So
unlike Eq. (\ref{E:VAC}), we are no longer
restricted to a local Lorentz frame and can
retain  Eq. (\ref{E:radial}).   As a
result, we can use Eqs. (\ref{E:rho}) and
(\ref{E:tangential}) to determine
\begin{equation}
   p_{\bot}(r)=-\rho(r)-\frac{r}{2}
   \frac{\partial\rho(r)}{\partial r}
   =p_r(r)
   +\frac{2\mu r^2\sqrt{\beta}}
   {\pi^2(r^2+\beta)^3}.
\end{equation}
So for larger $r$, we have
\begin{equation}
  p_{\bot}(r)\approx p_r(r).
\end{equation}
We can therefore
take the equation of state to be
\begin{equation}\label{E:isotropic}
   p(r)=-\rho(r)
\end{equation}
since the pressure is isotropic.

\section{The accelerating expansion}

\subsection{Large scale}\label{S:large}
Because of its isotropic nature, Eq.
(\ref{E:isotropic}) fits into a more
general cosmological setting.  (We
will return to the local frame in the
next subsection.)  In this setting, we
are dealing with a barotropic equation
of state $p=\omega\rho$, where $\omega$
is a constant.  Now, by Eq. (\ref
{E:isotropic}), $\omega =-1$, which
corresponds to Einstein's cosmological
constant, possibly the best model for
dark energy.  These observations are
consistent with the accelerated
expansion $\overset{..}{a}(t)>0$ in
the Friedmann equation
\begin{equation}\label{E:Friedmann}
   \frac{\overset{..}{a}(t)}{a(t)}
   =-\frac{4\pi}{3}(\rho+3p).
\end{equation}
By Eq. (\ref{E:isotropic}), it now
follows directly that
\begin{equation}\label{E:positive}
   -\frac{4\pi}{3}(\rho-3\rho)>0.
\end{equation}

\subsection{Small scale}
While Eq. (\ref{E:positive}) deals with
a global scale, our starting point, Eq.
(\ref{E:VAC}), assumes a local frame.
In this subsection, we return to a
microscopic scale to show that the
results are consistent with those above.

First we recall that a quantum fluctuation
is the temporary appearance of energetic
particles out of empty space, as allowed
by the uncertainty principle.  Although
transient fluctuations, they exhibit
some of the characteristics of ordinary
particles.  So both Eqs. (\ref{E:rho})
and (\ref{E:mass}) can be applied. Also,
since the Universe is a 3-sphere, any
point can be chosen for the origin.

Consider now the following metric in
Schwarzschild coordinates \cite{MTW}:
\begin{equation}\label{E:line1}
ds^{2}= -e^{\Phi(r)}dt^{2}+
\frac{dr^2}{1-\frac{2M(r)}{r}}
+r^{2}(d\theta^{2}+\text{sin}^{2}
\theta\,d\phi^{2}).
\end{equation}
(We are using units in which $c=G=1$.)
Here $M(r)$ is obtained in Eq.
(\ref{E:mass}).  If we remain
sufficiently close to the origin,
we can assume that $\Phi(r)\approx
\text{constant}$.  The Einstein field
equations are
\begin{equation}\label{E:Einstein1}
   \rho(r)=\frac{2M'}{8\pi r^2},
\end{equation}
\begin{equation}\label{E:Einstein2}
   p_r(r)=\frac{1}{8\pi}\left[-\frac{2M}{r^3}
   +\frac{2\Phi'}{r}\left(1-\frac{2M}{r}
   \right)\right],
\end{equation}
and
\begin{equation}\label{E:Einstein3}
   p_t(r)=\frac{1}{8\pi}\left(1-\frac{2M}{r}
   \right)\left[\Phi''-\frac{2M'r-2M}{
   2r(r-2M)}\Phi'+(\Phi')^2+\frac{\Phi'}{r}
   -\frac{2M'r-2M}{2r^2(r-2M)}\right].
\end{equation}
The conservation law  $T^{\alpha}
_{\phantom{\beta r}\beta;\,\alpha}=0$
implies that only Eqs. (\ref{E:Einstein1})
and (\ref{E:Einstein2}) are actually needed.
Since $\Phi'(r)\approx 0$, we now get
\begin{equation}\label{E:E1}
   \rho(r)=\frac{2M'(r)}{8\pi r^2}
\end{equation}
and
\begin{equation}\label{E:E2}
   p_r(r)=-\frac{1}{8\pi}\frac{2M(r)}{r^3}.
\end{equation}

Retaining our assumption of isotropic
pressure, Eqs. (\ref{E:E1}) and (\ref
{E:E2}) yield, in view of Eqs.
(\ref{E:rho}) and (\ref{E:mass}),
\begin{multline}\label{E:dark}
   \rho +3p=  \frac{\mu\sqrt{\beta}}
   {\pi^2(r^2+\beta)^2}-\frac{3}{8\pi}
   \left(\frac{1}{r^3}\right)
   \frac{4M}{\pi}\left(\text{tan}^{-1}
  \frac{r}{\sqrt{\beta}}-
  \frac{r\sqrt{\beta}}{r^2+\beta}
  \right)\\=
  \frac{1}{\pi^2\beta^{3/2}}
  \left[\frac{\mu}{(r^2/\beta+1)^2}
  -\frac{3}{2}\frac{M}
  {\left(r/\sqrt{\beta}\right)^3}
  \left(\text{tan}^{-1}
  \frac{r}{\sqrt{\beta}}-
  \frac{r/\sqrt{\beta}}{r^2/\beta +1}
  \right)\right].
\end{multline}
$M$ is likely to be only slightly
larger than $\mu$, making $\rho +3p$ only
slightly less than zero at the origin.
Fig. 1 shows the expression for $\rho +3p$
plotted against $r/\sqrt{\beta}$;
\begin{figure}[tbp]
\begin{center}
\includegraphics[width=0.8\textwidth]{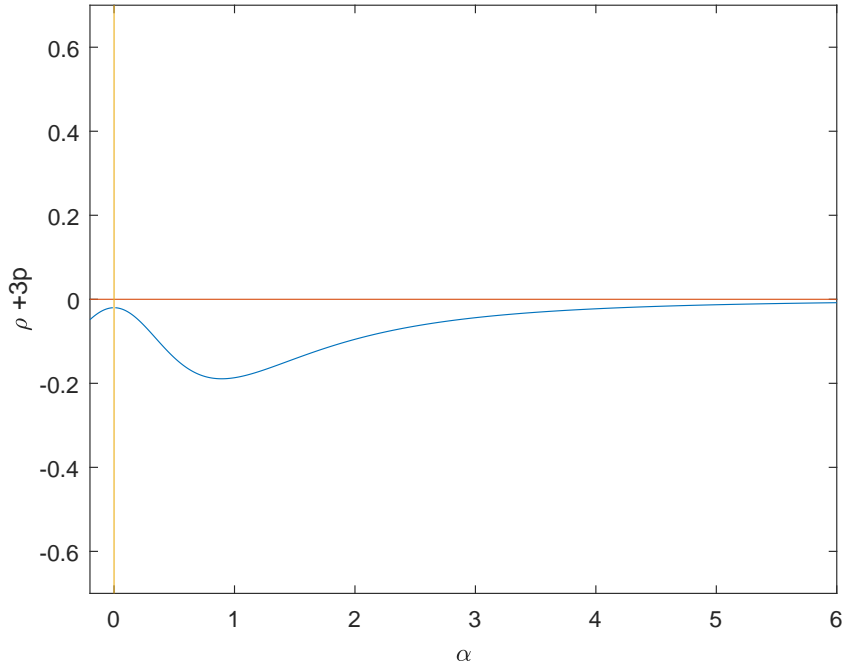}
 \end{center}
\caption{$\rho +3p$ is plotted against
   $\alpha =\frac{r}{\sqrt{\beta}}$.}
\end{figure}
so near the origin ($r$ close to zero and
$\sqrt{\beta}>0$), $\rho +3p<0$, a
result that is consistent with Eqs.
(\ref{E:Friedmann}) and (\ref{E:positive}).
So both the small-scale and large-scale
viewpoints lead to $\overset{..}{a}(t)>0$,
characteristic of dark energy, all
attributable to the noncommutative-geometry
background.

In Fig. 1, $\rho +3p$ is negative near the
origin and approaches zero asymptotically.
It is interesting to note that if the
smearing disappears altogether, then we
return to the classical setting: Fig. 1
shows that if $\sqrt{\beta}\rightarrow 0$,
then $\rho +3p<0$ for all $r$, which takes
us back to Subsection \ref{S:large}.  So
the local and global viewpoints complement
each other.  [That  $\text{lim}_{\sqrt{\beta}
\rightarrow 0}(\rho +3p)<0$ also follows
from Eq. (\ref{E:dark}).]

\section{Discussion}
It is proposed in Ref. \cite{gR20} that
vacuum energy is zero, as a result of
which the origin of dark energy must be
unrelated to vacuum energy.  The equation
of state $p_{\text{vac}}=-\rho_{\text{vac}}$
is a direct consequence of relativistic
invariance, according to Ref. \cite{gR20}.
It is also a consequence of a
noncommutative-geometry background.  The
latter does not require the local
Lorentz frame of the former.  The result
is that on both the local and cosmological
scales, $\rho +3p<0$, yielding
$\overset{..}{a}(t)>0$ in the Friedmann
equation.  A noncommutative-geometry
background, as formulated by Nicolini et
al., can therefore account for the
accelerated expansion and hence for dark
energy.  So vacuum energy can only be
zero if dark energy is indeed unrelated
to vacuum energy.

\end{document}